\documentclass{soups}
\usepackage{graphicx}
\usepackage{url}
\newcommand{\dc}{{DC}}
\newcommand{\uc}{{UC}}

\makeatletter
\let\@copyrightspace\relax
\makeatother 

\begin{document}
%
% --- Author Metadata here ---
%\conferenceinfo{Symposium on Usable Privacy and Security
%  (SOUPS)}{2010, July 14--16, 2010, Redmond, WA USA}
%\CopyrightYear{2010} % Allows default copyright year (200X) to be over-ridden - IF NEED BE.
%\crdata{0-12345-67-8/90/01}  % Allows default copyright data (0-89791-88-6/97/05) to be over-ridden - IF NEED BE.
% --- End of Author Metadata ---

\title{Security Through Entertainment: Experiences Using a Memory Game for Secure Device Pairing}

%\title{Alternate {\ttlit ACM} SIG Proceedings Paper in LaTeX
%Format\titlenote{(Produces the permission block, and
%copyright information). For use with
%SIG-ALTERNATE.CLS. Supported by ACM.}}
%\subtitle{[Extended Abstract]
%\titlenote{A full version of this paper is available as
%\textit{Author's Guide to Preparing ACM SIG Proceedings Using
%\LaTeX$2_\epsilon$\ and BibTeX} at

\numberofauthors{3} %  in this sample file, there are a *total*
\author{
% 1st. author
\alignauthor
Alexander Gallego\\
       \affaddr{Polytechnic Institute of NYU}\\
       \affaddr{Six MetroTech Center}\\
       \affaddr{Brooklyn, NY 11201}\\
       \email{\texttt{agalle01@students.poly.edu}}
% 2nd. author
\alignauthor
Nitesh Saxena\\
       \affaddr{Polytechnic Institute of NYU}\\
       \affaddr{Six MetroTech Center}\\
       \affaddr{Brooklyn, NY 11201}\\
       \email{\texttt{nsaxena@poly.edu}}
% 3rd. author
\alignauthor
Jonathan Voris\\
       \affaddr{Polytechnic Institute of NYU}\\
       \affaddr{Six MetroTech Center}\\
       \affaddr{Brooklyn, NY 11201}\\
       \email{\texttt{jvoris@isis.poly.edu}}
}
% 4th. author
%\alignauthor Lawrence P. Leipuner\\
%       \affaddr{Brookhaven Laboratories}\\
%       \affaddr{Brookhaven National Lab}\\
%       \affaddr{P.O. Box 5000}\\
%       \email{lleipuner@researchlabs.org}
% 5th. author
%\alignauthor Sean Fogarty\\
%       \affaddr{NASA Ames Research Center}\\
%       \affaddr{Moffett Field}\\
%       \affaddr{California 94035}\\
%       \email{fogartys@amesres.org}
% 6th. author
%\alignauthor Charles Palmer\\
%       \affaddr{Palmer Research Laboratories}\\
%       \affaddr{8600 Datapoint Drive}\\
%       \affaddr{San Antonio, Texas 78229}\\
%       \email{cpalmer@prl.com}
%}
% There's nothing stopping you putting the seventh, eighth, etc.
% author on the opening page (as the 'third row') but we ask,
% for aesthetic reasons that you place these 'additional authors'
% in the \additional authors block, viz.
%\additionalauthors{Additional authors: John Smith (The Th{\o}rv{\"a}ld Group,
%email: {\texttt{jsmith@affiliation.org}}) and Julius P.~Kumquat
%(The Kumquat Consortium, email: {\texttt{jpkumquat@consortium.net}}).}
% Just remember to make sure that the TOTAL number of authors
% is the number that will appear on the first page PLUS the
% number that will appear in the \additionalauthors section.

\maketitle
\begin{abstract}

The secure ``pairing'' of wireless devices based on auxiliary or out-of-band (OOB) communication, such as audio, visual, or tactile channels, is a well-established research direction. However, prior work shows that this approach to pairing can be prone to human errors of different forms that may directly or indirectly translate into man-in-the-middle attacks. To address this problem, we propose a general direction of the use of computer games for pairing. Since games are a popular means of entertainment, our hypothesis is that they may serve as an incentive to users and make the pairing process enjoyable for them, thus improving the usability, as well as the security, of the pairing process.

We consider an emerging use case of pairing whereby two different users are
involved, each in possession of his or her own device (e.g., Alice and Bob
pairing their smartphones for social interactions). We develop ``Alice Says,''
a pairing game based on a popular memory game called Simon (Says), and discuss
the underlying design challenges.  We also present a \textit{preliminary}
evaluation of Alice Says via a usability study and demonstrate its feasibility
in terms of usability and security. Our results indicate that overall Alice
Says was deemed as a fun and an enjoyable way to pair devices, confirming our
hypothesis. However, contrary to our intuition, the relatively slower speed of
Alice Says pairing was found to be a cause of concern and prompts the need for
the design of faster pairing games. We put forth several ways in which this
issue can be ameliorated. In addition, we also discuss several other security
problems which are lacking optimal solutions and suggest ideas on how
entertainment can be used to improve the current state of the art solutions
that have been developed to address them. 

\end{abstract}

%A category including the fourth, optional field follows...
%\category{D.2.8}{Software Engineering}{Metrics}[complexity measures, performance measures]

% Your general terms must be any of the following 16 designated terms:
% Algorithms, Management, Measurement, Documentation, Performance,
% Design, Economics, Reliability, Experimentation, Security, Human
% Factors, Standardization, Languages, Theory, Legal Aspects, Verification

% You are on your own with keywords

\section{Introduction}

\label{sec:intro}

Short and medium-range wireless communication based on technologies such as
Bluetooth, WiFi, and RFID (Radio Frequency IDentification), is becoming
increasingly popular and promises to remain so in the future.  This surge in
popularity unfortunately brings various security risks along with it. Wireless
communication channels are easy to eavesdrop upon and manipulate. Therefore, a
fundamental security objective is to secure such data transfer mediums. In this
paper, we use the term ``pairing'' to refer to the operation of bootstrapping
secure communication between two wireless devices in a way that is resistant to
eavesdropping and man-in-the-middle attacks. Examples of common use cases for
this operation include pairing between a headset and phone, or between two
smartphones. The initialization of secure communication would be easy to
achieve if there existed a global infrastructure enabling devices to share an
on or off-line trusted third party, certification authority, PKI or
pre-configured secrets.  However, such a global infrastructure may not be
possible in practice, thereby making pairing an interesting and a challenging
research problem.

% The problem has been at the forefront
%of various recent standardization activities; see \cite{SVA07}.  \cite{wifi},
%\cite{net}, \cite{BT} and \cite{USB}.}

A promising and well-established research direction to solving the pairing dilemma is to leverage an auxiliary channel, also called an out-of-band (OOB) channel, which is governed by the users operating the devices to be paired. Examples of OOB channels include audio, visual, and tactile channels. Unlike classical radio channels, OOB channels are ``human-perceptible,'' i.e., the underlying transmission and reception that drives these avenues of communication can be perceived by one or more of human senses. Due to this property, OOB communication naturally provides authentication and integrity, unlike radio communication. In other words, a user can validate the intended source of an OOB message and an adversary can not manipulate the OOB messages in transit, although he can perform a variety of other actions, such as eavesdropping upon data sent across the channel.  

%The earlier pairing protocols \cite{Balfanz02} \cite{PV06-1} require at least
%$80$ to $160$ bits of data to be transmitted over the A-OOB channels.

%\footnote{The concept of SAS-based authentication was first introduced by
%Cagalj et al. \cite{CCH06}, followed by Vaudenay \cite{Vaudenay05}. MANA
%protocols \cite{MANA} addressed a similar problem.}

The usability of pairing based on OOB channels is clearly very important. Since the OOB channels typically have low bandwidth, the shorter the data that a pairing method needs to transmit over these channels, the better the method becomes in terms of usability. To this end, a recent innovation in pairing are the so-called Short Authenticated String (SAS) based protocols \cite{Nyberg05,PV06,CCH06,MANA,Vaudenay05} that limit the length of data to be transmitted over OOB channels to only $15$ bits or so, while achieving a reasonable level of security. Using these protocols, a wide-variety of pairing methods based on visual, audio, tactile, and infrared OOB channels have been proposed. We refer the reader to a survey and comparative analysis of various OOB pairing methods \cite{kstu09}. (We will later summarize these in Section \ref{sec:related_pairing}).

The focus of this paper is on \textit{social pairing} scenarios \cite{2-user-tech}, whereby two different users (Alice and Bob) control their respective devices while pairing them. Examples include pairing between Alice's and Bob's PDAs, laptops, or cell phones for social or professional reasons, such as sharing files and music, exchanging digital business cards, multiplayer games, messaging, chatting, or collaborative applications.  The main advantage of using Bluetooth or WiFi in such scenarios is that no infrastructure is needed and thus \textit{ad hoc} communication can take place without any extra cost to the users. For this reason, social scenarios have been emerging rapidly and are already quite popular, especially in developing countries. Secure pairing of users' devices is a natural and recommended way to prevent any eavesdropping and/or malicious intervention during their intended communication. Furthermore, note that most scenarios necessitating OOB pairing techniques are by definition social in nature. This is because of the fact that if a single user is the administrator of both of the devices to be paired, he or she can simply use a pre-shared secret on both devices to accomplish pairing in a straightforward fashion.

%\textit{constrained devices}. We define a
%constrained device as a device that lacks good quality output interfaces (e.g,
%a speaker, display), input interfaces (e.g., keypads),  or receivers (e.g.,
%microphone, camera), and may not be physically accessible.  Examples of
%constrained devices include headsets, access points, and medical
%implants.\footnote{Due to economic reasons, such devices may also be
%constrained in terms of computational resources (e.g., low-cost RFID tags).}
%Establishing OOB channels on constrained devices, therefore, might be quite
%difficult. 

\subsection{Research Challenges}
\label{sec:challenges}

We remark that the problem of social pairing is simpler than a commonplace problem of \textit{personal pairing}, whereby both devices are controlled by a single user (Alice). Examples of personal pairing include pairing between Alice's Bluetooth headset and her cell phone, her PDA and her wireless printer, or her laptop and a wireless access point. This is because, unlike personal pairing, the devices taking part in social pairing are not usually constrained in terms of input/output interfaces. In fact, most modern cell phone class of devices are equipped with a wide variety of interfaces which make establishment of OOB channels much simpler. 

Unfortunately, even the seemingly simple problem of social pairing turns out to be daunting in practice and remains unsolved despite being subject to several recent years of research. Prior work on pairing raises several usability and security related concerns and fundamental research challenges. The most prominent of these challenges are as follows:

\begin{itemize}

\item Most existing pairing methods are based on SAS protocols that use very short strings, perhaps only 15 bits in length. The level of security provided by these methods may therefore not be sufficient for certain applications. Increasing the length of SAS strings, on the other hand, may lead to poor usability and security because the process will become lengthier.  Methods that are automated (e.g., based on cameras) and can transmit longer SAS strings, are also shown to have undesirable usability properties \cite{kstu09}.

\item Even while using short OOB strings, several comparison-based pairing methods (i.e., those based on comparison of OOB strings) do not offer the theoretical level of security guaranteed by their underlying protocols, as demonstrated in \cite{kstu09}. This is due to the potential these protocols have for human errors. Such errors can be of two forms: \textit{fatal} and \textit{safe} \cite{uka06}. Fatal errors occur when a user accepts a pairing instance, although the OOB strings on the two devices did not match, leading to the potential for a man-in-the-middle attack.  Safe errors, on the other hand, occur when a user rejects a pairing instance even when the OOB strings on the two devices match. Such errors undermine the usability of pairing, but can also have an indirect impact on security; a failed pairing necessitates repetition, which may lead to user annoyance and translate into attacks eventually.

\item A more serious issue is that security of pairing often has to rely upon
the decision made by the users. As a result, a \textit{rushing user} \cite{su09}\footnote{A
rushing user is a user who -- in a rush to connect her devices -- would skip
through the pairing process, if possible \cite{su09}.} may simply just
``accept'' the pairing, without having to correctly take part in the decision
process. Pairing methods that are based on transfer of OOB strings (and
decision made by the devices instead) are naturally resistant to rushing user
behavior, but are still prone to safe errors \cite{uka06}. 

\end{itemize}

The aforementioned challenges motivate the design of a radically different approach to pairing. The central research question can be summarized as: {can we design pairing methods that can handle longer OOB strings, and are as resistant as possible to potential safe and fatal errors, as well as to the rushing user behavior?} This question can, alternatively and fundamentally, be framed as follows: \textit{can we design pairing methods that incentivize the users, in some way, so that they correctly take part in the pairing process, thus providing improved security as well as user experience?}
%\end{itemize}

\subsection{Motivation: Games for Pairing}
\label{sec:motivation}

To help answer the above question, we propose a general direction of the
application of computer games for pairing of devices. The incentive that we
provide to the user, while they pair their devices, is fun and entertainment.
Since games are a popular means of entertainment, our hypothesis is that they
may improve the security as well as usability of pairing, and help solve the
challenges outline above. 

%tom sawyer effect

We try to delve deeper as to \textit{why a game should be used to address the
problem of device pairing}. If the pairing of personal wireless devices was
known to be a solved problem with a straightforward satisfactory solution, this
would not be necessary. This is not the case, however, as discussed in Section \ref{sec:challenges}. 
%While many pairing
%solutions have been proposed, some of which are outlined in Section
%\ref{sec:related_pairing}, 
%many of these solutions necessitate that user
%involvement in mundane tasks, such as closely monitoring patterns of visual,
%auditory, or tactile feedback. 
%As alluded to in Section
%\ref{sec:related_games}, 
Users may not be aware of or care about the impact
their actions during this process may have on the security of their appliances.
As a result of this lack of engagement in the process of pairing, they may not
do their best to complete the pairing or may attempt to skip it entirely, if possible.

To address this issue, we propose the reframing of the pairing process not as a
tedious procedure that puts a costly burden on users, but rather as a game that
is enjoyable and entertaining to complete. It is our aim to transform the
operation of device pairing from one that users seek to avoid or complete as
quickly as possible into one that they relish. As a result, users will be more
attentive to the steps they must follow while pairing and perform better at
it. Furthermore, if a game involves competitiveness between more than one
individual, this will provide another layer of motivation to put forth their
best possible performance. Another important side effect of a game can be that,
due to its entertainment value, users will be willing to spend more time during
the pairing process. This way potentially longer OOB strings can be used, thus
providing higher level of security.

In essence, we are suggesting that by contextualizing a security task as a game
rather than a chore, the usability burden of this task can be greatly reduced.
We dub this the \textit{Tom Sawyer Effect} after a well known event in Mark Twain's
literary classic, ``The Adventures of Tom Sawyer'' \cite{tomsawyer}. In this
novel, the boy Tom Sawyer is chastised by his Aunt Polly by being forced to
paint a fence on his day off. Tom resents the fact that he must complete this
task instead of enjoying his free day by playing games with his friends. To
escape his plight, the clever Sawyer acts as though he is having a good time
performing his task rather than resenting it. Upon observing his supposed
delight, his friends insist that they be given an opportunity to paint the
fence so that they can enjoy it as well, going as far as to trade him trinkets
for the opportunity to do so. Much in the same way that Tom convinces his
friends to complete what would otherwise be considered an uninteresting job by
treating it as a game, we seek to persuade users to be attentive during
security operations, such as device pairing, by making them as entertaining as
possible.

\subsection{Contributions}

We develop ``Alice Says,'' a pairing game based on a popular memory game
called Simon (Says), and discuss design challenges behind this
construction.
%The steps that went into its implementation are outlined
%as well.
We also present a \textit{preliminary} evaluation of Alice Says
via a usability study and demonstrate its feasibility in terms of
usability and security. Our results indicate that, overall, Alice Says
was deemed to be a fun and enjoyable way to pair devices, confirming our
hypothesis. It was also found to be robust to human mistakes. However,
contrary to our intuition, the speed of Alice Says pairing was found to
be a cause of concern and prompts the need for the design of faster
pairing games. We put forth several ways in which this issue can be
ameliorated. In addition, we also discuss several other security
problems which are lacking optimal solutions and suggest ideas on
how entertainment can be used to improve the current state of the art
solutions that have been developed to address them.

\medskip
\noindent \textit{Outline:} 

The remainder of this paper is organized as follows. First, in Section
\ref{sec:related}, we discuss prior device pairing methods and the work that
relates to the use of games for security applications. In Sections
\ref{sec:design} and \ref{sec:implementation}, we present the design and
implementation of our pairing game Alice Says. This is followed by Section
\ref{sec:experimentation} where we report our first usability evaluation of
Alice Says. Finally, in Section \ref{sec:discussion}, we discuss the results and
implications of our study, the lessons learned and possible applications of
games to solve other security problems.

\section{Related Work}

\label{sec:related}

\subsection{Prior Pairing Methods}
\label{sec:related_pairing}

In this subsection, we discuss prior pairing methods, and, in particular, 
outline whether or not they are
resistant to rushing user behavior. In doing so, we distinguish the methods
into two categories (as discussed in  \cite{su09}): device-controlled (\dc) and user-controlled (\uc)
following the terminology introduced in \cite{su09}. In a \dc\ method, device
decides the outcome of paring, whereas in a \uc\ method, user decides the
outcome of pairing. Note that a \dc\ method would be naturally resistant to
rushing user behavior, but a \uc\ method is not.

In their seminal work, Stajano, et al.\ \cite{SA99} proposed establishing a
shared secret between two devices using a link created through a physical
contact (such as an electric cable). This is a \dc\ method and is resistant to
rushing user. However, in many settings, establishing such a physical
contact might not be possible, for example, the devices might not have common
interfaces to do so or it might be too cumbersome to carry the cables along.
Balfanz, et al. \cite{Balfanz02} extended this approach through the use of
infrared channel -- the devices exchange their public keys over the
wireless channel followed by exchanging (at least $80$-bit long) hashes of
their respective public keys over infrared. This is also a \dc\ method. The main
drawback of this method, however, is that it is only applicable to devices
equipped with infrared transceivers. Moreover, the infra-red channel is not
easily perceptible by human users.

Another approach taken by a few research papers is to perform the key exchange
over the wireless channel and authenticate it by requiring the users to
manually and visually compare the established secret on both devices. Since
manually comparing the established secret or its hash is cumbersome for the
users, methods were designed to make this visualization simpler. These include
Snowflake mechanism \cite{Gold96} by Levienet et al., Random Arts visual hash
\cite{PS99} by Perrig et al., etc. These methods require high-resolution
displays and are thus only applicable to a limited number of devices, such as
laptops. Moreover, these are \uc\ methods and thus are vulnerable to a rushing
user. 

Based on the pairing protocol of Balfanz et al. \cite{Balfanz02}, McCune et al.
proposed  the ``Seeing-is-Believing'' (SiB) method \cite{MPR05}. SiB involves
establishing two unidirectional visual channels -- one device encodes
the data into a two-dimensional barcode and the other device reads it using a
photo camera. SiB is a \dc\ method. However, since the method requires both
devices to have cameras, it is only suitable for pairing devices such as camera
phones. Moreover, a recent study \cite{kstu09} shows that users may not be
comfortable handling cameras and this method may not be very usable. 

Goodrich, et al. \cite{lac05}, proposed ``Loud-and-Clear (L\&C)'', a pairing
method based on ``MadLib'' sentences. The main idea of L\&C is to encode the
OOB data into MadLib sentences and have the user compare these sentences
displayed or spoken out on two devices. Clearly, this is a \uc\ method and is
thus vulnerable to rushing user behavior. Moreover, the method is not
applicable to pairing scenarios where one of the devices does not have a
display or a speaker. 

Saxena et al. \cite{seka06} proposed a pairing method based on visual OOB channel.
The method uses one of the SAS protocols \cite{Nyberg05}, and is aimed at
pairing two devices A and B (such as a cell phone and an access point), only
one of which (say, B) has a relevant receiver (such as a camera).  First, a
unidirectional channel is established by device $A$ transmitting the SAS
data, e.g., by using a blinking LED and device $B$ receiving it using a video
camera. This is followed by device $B$ comparing the received data with its own
copy of the SAS data and displaying the result of comparison.  Finally, the user reads the result and accordingly indicates the result to device $A$.  In one direction (i.e., from
A to B), this is a \dc\ method and is thus resistant to rushing user behavior.
In the other direction, however, it is a \uc\ method -- a rushing user can
simply accept the pairing on A without looking at the pairing outcome on B.  
%It
%is important to note, however, that in case of any attack, device B will be
%``locked out'' and will not allow any connection to and from device A (and it
%will detect any connection attempts from an attacking device).\footnote{In case of a pairing failure, device B
%can keep showing a warning to the user indicating that device A is
%possibly being connected to an attacker device, and ask the user to ``re-pair''
%the two devices.} This way the user will not be able to establish real
%communication with device B (e.g., transfer an image file from B to A), and
%will thus resort to repeating the pairing process. The pairing methods we
%propose in this paper utilize this unidirectional pairing approach of
%\cite{seka06}. 

Uzun et al.\ \cite{uka06} carry out a comparative usability study of simple
pairing methods. They consider pairing scenarios where devices are capable of
displaying $4$-digits of SAS data. In what they call the
``Compare-and-Confirm'' approach (a \uc\ method), the user simply reads and
compares the SAS data displayed on both devices.  The ``Select-and-Confirm''
approach (a \dc\ method), on the other hand, requires the user to select a
$4$-digit string (out of a number of strings) on one device that matches with
the $4$-digit string on the other device. The third approach, called
``Copy-and-Confirm'' (a \dc\ method), requires the user to read the data from
one device and input it onto the other.  Both Select-and-Confirm and
Copy-and-Confirm are \dc\ methods. However, since they are based on 
the protocol of \cite{seka06}, they offer protection against a
rushing user only in one direction.   
%However, these methods are only limited to devices (such as cell
%phones) which have good quality displays and keypads. Our alphanumeric pairing
%method is quite similar in flavor to the Copy-and-Confirm method of
%\cite{uka06}, however, it is applicable to interface-constrained devices. 
Kuo et al.\ \cite{Kuo07} defined a common baseline for hardware features and 
a consistent, interoperable user experience across pairing of different
 devices. This work did not yield any pairing method as such.

Some recent papers have focused upon pairing devices which possess constrained
interfaces, including access points, headsets, which lack
good quality output interfaces (e.g., a speaker, display) and/or receivers
(e.g., microphone, camera). These include the BEDA method \cite{beda} which
requires the users to transfer the SAS strings from one device to the other
using ``button presses''. In \cite{sr08,wisec}, Saxena et al. presented similar
pairing methods universally applicable to any pair of devices. The method can
be based on any of the existing SAS protocols and does not require devices to
have good transmitters or any receivers, that is, just a pair of LEDs is
sufficient.  These method involves users comparing very simple audiovisual
patterns, such as ``beeping'' and ``blinking''.  Most recently, the approach of
\cite{sr08} was extended by making use of an auxiliary device, such as a
smartphone \cite{suv08}. Both these methods, however, are \uc\ methods and thus
offer no protection against a rushing user.
 
In \cite{hapadep}, Soriente et al.\
consider the problem of pairing two devices which might not share any common
wireless communication channel at the time of pairing, but do share only a
common audio channel. This is a \dc\ method, however, it is only limited to
devices which possess a speaker at the transmitting end and a microphone at the
receiving end. Moreover, this method still requires the user to perform manual
comparison of SAS data (e.g., using the L\&C method \cite{lac05}) and is thus
not resistant to rushing user behavior. 

\subsection{Games and Security}
\label{sec:games-security}
%games for pairing paper

Our work was in a way inspired by the work of Halprin and Naor \cite{game_rng}.
These researchers recently proposed the use of games to address the problem of
computer random number generation. Computers often use inputs from users as an
entropy source.  Unfortunately, when asked to cooperate in this endeavor, human
users tend to perform poorly by interacting with the machine in a predictable
fashion. 
%There are several reasons behind this. One is that generating entropy through
%methods such as striking random keys or moving a mouse is simply not an
%engaging process, and users lack any motivation to perform well at it.
%Furthermore, users might not understand what a ``good'' way to input
%randomness is or why it is important, particularly those who are unaware of
%the security consequences, which the software requesting the randomness often
%times does not make clear. 

More critically, human beings are notoriously bad at behaving randomly or
recognizing randomness in a natural setting. When asked to construct random
sequences, people's output are riddled with numerous biases. There remains some
hope for human entropy generation, however. Interestingly, when placed in
competitive situations \cite{game_rng}, such as zero sum games or any other situation where
individuals are asked to attempt to outperform one another, humans demonstrate
a heightened aptitude for behaving randomly. Therefore, security gains are
noted over traditional entropy generation requests when users are asked to
participate in a game that forces them to behave randomly and then harvests
entropy from their actions.

The pairing mechanism that we present in this work is an example of a ``Game with a Purpose'' as conceptualized by von Ahn \cite{gwap}. This is because it is not simply a game for its own sake, but rather a form of entertainment that simultaneously achieves a computational result without the explicit awareness of its users. The reCAPTCHA \cite{reCAPTCHA} project of von Ahn et al. is also tangentially related to this line of research. While it does not involve any games or entertainment, this program also ``fools'' users into doing work beyond what they may realize. A CAPTCHA is a tool that attempts to distinguish between a human user and a computerized entity by presenting requesters with a task that is relatively easy for a human to accomplish but more difficult for a computer to achieve. This most often takes the form of presenting users with a word or phrase that has been obscured, distorted, or is otherwise difficult to read. reCAPTCHA not only serves this adjudicative purpose, but also utilizes the responses it receives to aid in the digitization of words in print media.

\section{Design of a Pairing Game}
\label{sec:design}

\subsection{Threat Model}
\label{sec:threat_model}

Before discussing the design of our secure device pairing game, Alice Says, it
is first necessary to establish the adversarial model for the scenario it is
intended to operate in. (This is the same model followed by various pairing
methods based on the SAS protocols \cite{Vaudenay05}.) The two wireless devices
to be paired may establish two types of communication channels between each
other. The first is a traditional wireless connection. This type of channel is
characterized by a large bandwidth capacity and bidirectionality, but is short
range in nature. The second variety compromise the set of OOB channels, which
feature relatively modest bandwidths but are physically authenticatable. That
is, OOB channels are crafted from forms of output which can be perceived by
unassisted humans, providing users with the power to verify the transmission
source themselves.  This implies that any malicious entity who wishes to
manipulate data in transit over an OOB channel would not be capable of
modifying any messages. OOB channels are not generally secret, however.  This
means that besides the modification restriction, adversaries can observe the OOB
transmission in any other way they see fit. 
%This includes listening in on the contents of the channel as well as deleting
%messages, replaying them, or reordering them. 
In contrast, opponents have unfettered control of the conventional wireless
channel, and can operate on messages sent across it as they see fit. 

\subsection{Choice of the Game}

%established game
%already known to be popular
%ease of implementation
%easy to play for people who might not be considered computer "gamers"
%very similar to previous methods of solving the pairing problem using synchronized forms of human percievable outputs
%web sites for Simon: http://www.thepcmanwebsite.com/media/simon/ http://en.wikipedia.org/wiki/Simon_(game)

In order to leverage the Tom Sawyer effect to improve the device pairing
experience (as discussed in Section \ref{sec:motivation}), a suitable game had
to be designed. We took our inspiration from an existing game, Hasbro's Simon
\cite{simon}. While this game was originally a freestanding electronic device,
many derivatives have been created that can be played on mobile devices or
through a web browser on a traditional computer, such as the one found here
\cite{simon_web}. 

This game was selected as a basis for our pairing game for several reasons.
First and foremost, Simon is a well established game, having been created over
three decades ago, with a loyal following. Rather than creating a new game from
scratch, which users may or may not find enjoyable, we hoped to leverage the
known popularity of Simon.  Furthermore, this game is relatively uncomplicated
when compared with the contemporary generation of computer games. This was
desirable both due to its ease of implementation, which did not require a large
team of programmers well versed in graphical and game programming, as well as
its suitability for players of all ages and levels of experience. That is,
Simon was selected to appeal to as broad a swath of users as possible rather
than a niche group of die hard computer ``gamers.'' Finally, an important
factor in the selection of this game is its close relation to existing device
pairing solutions. Previous work has established the use of patterns of
synchronized audio and visual output \cite{sr08,wisec,beda} as a viable method
of securely associating devices. At its core, playing Simon involves
nothing more than the short term memorization of audiovisual patterns and thus
minimal changes were required to adapt it for use in pairing.

\subsection{Alice Says Game Design}

In order to answer the questions put forth in Section \ref{sec:intro},
specifically, the effect of entertainment on the device pairing process, we
developed a pairing game dubbed Alice Says based on the classic electronic
memory game, Simon (Says). 
%While the intent of Alice Says is to securely link personal wireless devices,
%care was taken to avoid any reference to the term ``pairing'' or anything else
%related to security in the game itself.  This was critical to prevent the
%explicit priming of test subjects. 
Upon initially starting the game, users are provided a screen showing the name
of the game with two menu choices: a single player training mode and a two
player pairing mode. Optionally, an all-time high score can also be displayed.

\subsubsection{Single Player Mode}

In accordance with the Tom Sawyer effect we crafted Alice Says to promote, a
single player mode is provided to allow users an opportunity to unwittingly
train themselves to improve their device pairing performance. This operational mode is
essentially identical to the classic version of Simon, only adapted to the
context of a mobile device. The user is shown a screen with four adjacent
squares which fully occupy the screen, dividing it into quadrants. Each of
these squares is a unique and distinctive color. Clockwise starting in the
upper left, the colors are green, red, blue, and finally yellow. Besides these
color coded screen segments, the only other item visible to the user while the
game is underway is a counter which tracks the length of the pattern that a
user has matched thus far. This increments with each new button press that is
added to the pattern list.

One of these four quadrant buttons is randomly selected by the device during
each round. The selection is indicated to the user in two complimentary ways.
First, the screen section is lit by increasing its luminance. Secondly, a tone
corresponding to that quadrant is played. In order to maximize the enjoyment of
the game by users, the notes associated with the four portions of the screen
are carefully selected to be harmonically compatible with each other
irrespective of the order in which they are played. Indeed, this was a critical
component behind the game's widespread popularity. Since the Simon user
interface consists of four color quadrants, each step of the pattern can be
used to encode two OOB bits in the following straightforward manner:
``00'' corresponds to green, ``01'' is indicative of red, ``10'' means blue,
and ``11'' is aligned with yellow.

If a user presses the screen quadrant that correctly corresponds to the one
that had just been selected by the device, it then constructs a pattern of
colors and sounds by displaying the first screen segment followed by a new,
randomly chosen one, again conveyed to the user by brightening the relevant
portion of the screen and playing a corresponding melodic tone. This process
continues until the user makes an error in the pattern or a certain predetermined pattern length threshold value is reached. At this
point a ``Game Over'' message is provided to the user, informing him or her of
what the correct move should have been as well as the length of the pattern
they were successfully able to match. If a high score is maintained and was
surpassed, this is updated at this point as well.

\subsubsection{Two Player Mode}

%why two player? pairing devices, so two parties will always be present. also,
%competitive element should theoretically bolster performance by participants
The two player mode is what actually accomplishes device pairing. It differs
from the single player, traditional Simon approach in two main ways. First, the
game does not conclude when a mistake is made. It continues until a sufficient
number of OOB bits have been relayed between the two devices. Secondly, the
game is split across two devices. One device displays the pattern to the user,
but does not handle input. The other does not display the pattern, but instead
only accepts user input. A two player pairing game was selected in lieu of a
single player game because the need to pair two wireless devices implies that
there will always be two parties present to participate in the pairing. Less
intuitively, a two player game will also reap the security benefits of
fostering competition between its players, hopefully resulting in an
amplification of the Tom Sawyer effect.

As a result of the second difference, a mechanism is required to keep the two
devices in sync during the pairing procedure. This is because the device
accepting input must be aware of what portion of the pattern has been displayed
to the user to be able to conclude whether or not he or she has committed a
mistake. A naive way to handle this would be to simply transmit the current
index into the binary OOB string over the wireless channel. This would ruin
the security of the system, however, as the wireless channel is assumed to be
totally insecure as per the security assumptions detailed in Section
\ref{sec:threat_model}. Thus, an adversary could simply transmit an arbitrary
index over the string, bypassing as much of it as he or she desired. Instead,
we addressed the synchronization issue by integrating ``previous'' and ``next''
buttons into the displaying phone's interface. 

Upon a successful match, the player in control of the display phone presses
next to advance the state of the game and display the next, slightly longer
pattern. If an error is made while inputting the pattern, on the other hand,
this user can indicate that this is the case to the displaying device by
pressing the previous button. In this event, a new pattern is crafting starting
with the last bit of the pattern which was matched unsuccessfully. This is done
because, as an invariant, the devices know that the users have already
successfully transferred all OOB bits up until this point. After a mistake is
made, the input device will check its running tally of how many bits have been
successfully matched compared to how many bits of the OOB string need to be
transferred.If more bits need to be compared, the game continues. If mutual
authentication is desired, the roles of the phones can be swapped following a
successful game in a single direction. After this, the game play would proceed
in precisely the same way with the roles of the two phones, and two users,
reversed.

\subsubsection{Example Usage Scenario}

The following is an example of an anticipated game play pattern. Assume a
legitimate pairing session in which a user is consistently able to follow a
pattern of length 5. There are 30 bits in the OOB string that need to be
compared. The user will first be provided with a pattern of length one, then
the pattern will be extended to length two, and so on. On the sixth round, the
user will make a mistake. Having successfully exchanged the first 10 bits of
the key hash, the game will then begin a new pattern starting with the 11th and
12th bits of the OOB string. Upon making another mistake at length 6, the game will
begin with a new pattern starting with the 21st and 22nd key OOB bits. After
successfully completing the next session of 5 bits, all 30 bits will have been
conveyed, concluding the game.

\subsection{Security Guarantees}

One important, and subtly tricky, aspect to this design is how to handle
attacked sessions. If a session has been attacked or an error has occurred, the
OOB strings calculated on the two devices will be different. Thus, even if a
user ``correctly'' matches the displayed pattern, Alice Says will still
register an error. This will either occur as the first bit of a pattern or a
sub-sequential bit of the pattern. If the attack occurs as a sub-sequential bit,
the bits prior to the error will be registered as a match for that session and
the pattern will begin anew with the attacked bit, making it the first bit of
the next pattern. Thus, one way or another, the attacked bit will end up as the
first bit of a pattern, and users will be unable to proceed by identifying the
single color pattern that has been displayed to them.

Thus, users will need to be provided with a mechanism for restarting the
pairing session from scratch. To achieve this, after a certain threshold of
single color pattern mismatches have occurred (e.g., 2), a message along the
lines of ``It seems that something has gone wrong. Would you like to restart
the session? [Y/N]'' will be displayed. At this point, users can scrap the
entire session and start over with a new, hopefully unattacked and error free,
session. Note that single color pattern mismatches should be very unlikely in
unattacked sessions, as most users are anticipated to be able to handle
matching at least one color. Thus, the only way for a critical error to occur
in this system is for users to incorrectly match a single color. There is only
a 1 in 4 chance of this occurring, even if the user is not paying any attention
whatsoever and just guessing at the pattern, yielding a high level of practical
security. Note that while it is theoretically possible for a player to complete
pairing with Alice Says by making random color quadrant choices, this would
take prohibitively long to achieve, and it is therefore not plausible for such
an attack to succeed.

\section{Implementation of Alice Says}

\label{sec:implementation}

We set out to develop Alice Says, our game that would preserve the popular aspects of
the classic electronic Simon while updating it to a two player mobile device
setting. Its user interface is dominated by four large color buttons as was the
case with its ancestor. Also intended to mimic the original was the association
of a unique tone with each of these keys. A critical aspect of the original
game's appeal was the fact that these sounds were designed to harmonic
irrespective of the order in which they were played, which is important as the
game play involves striking the inputs in a random order. Thus we tried to stay
true to the original game's sounds by assigning an A note to the first input,
an A note one octave higher to the second, a D note that is a perfect fourth
above the initial A note to the third button, and finally a G note that was a
perfect fourth higher than the D to the last key.

\subsection{Java Micro Edition}

We utilized two Nokia model N97 mobile phones to realize our Alice Says prototype. These devices support the Java Platform, Micro Edition (J2ME) environment, which we designed our code to operate in. Since the primary objective of our device pairing prototype was to test its usability, we focused largely on crafting a user interface that was as intuitive and user friendly as possible. Thus, one of the most essential design decisions that needed to be made was which J2ME API to build our game upon. Initially we used the Lightweight User Interface Toolkit to this end, as it provided the most streamlined interface available. Unfortunately, the use of this API proved to be problematic. While the N97 model phones met the toolkit's requirements, this particular phone model's Java Virtual Machine implementation caused some of its components to crash at run time. This was discernable due to the fact that while library executed properly while being emulated but did not operate properly on the devices themselves.

As an alternative, we instead utilized the lower level APIs provided by the Nokia N97 SDK. The majority of the graphics used in Alice Says were drawn using the Graphics class of Java Specification Request (JSR) 118 in the 2.0 revision of the Mobile Information Device Profile. This  low level graphics object provided built in function calls that allowed us full control over painting the screen as well as handling user inputs through the N97's touch screens. Menu options were handled using JSR 118's CommandListener class. To further enhance our design, the device accepting user input had no commands available to go back or forward in the Alice Says pattern. Instead, it kept track of whether or not an error occurred and automatically adjusted the pattern length in accordance with user errors. On the other hand, the device displaying the pattern accepted no input to prevent the possibility of a user disrupting the pattern by mistake. 

\subsection{Device Setup}

\begin{figure}[htbp]
\centering{
  \includegraphics[width=.45\textwidth]{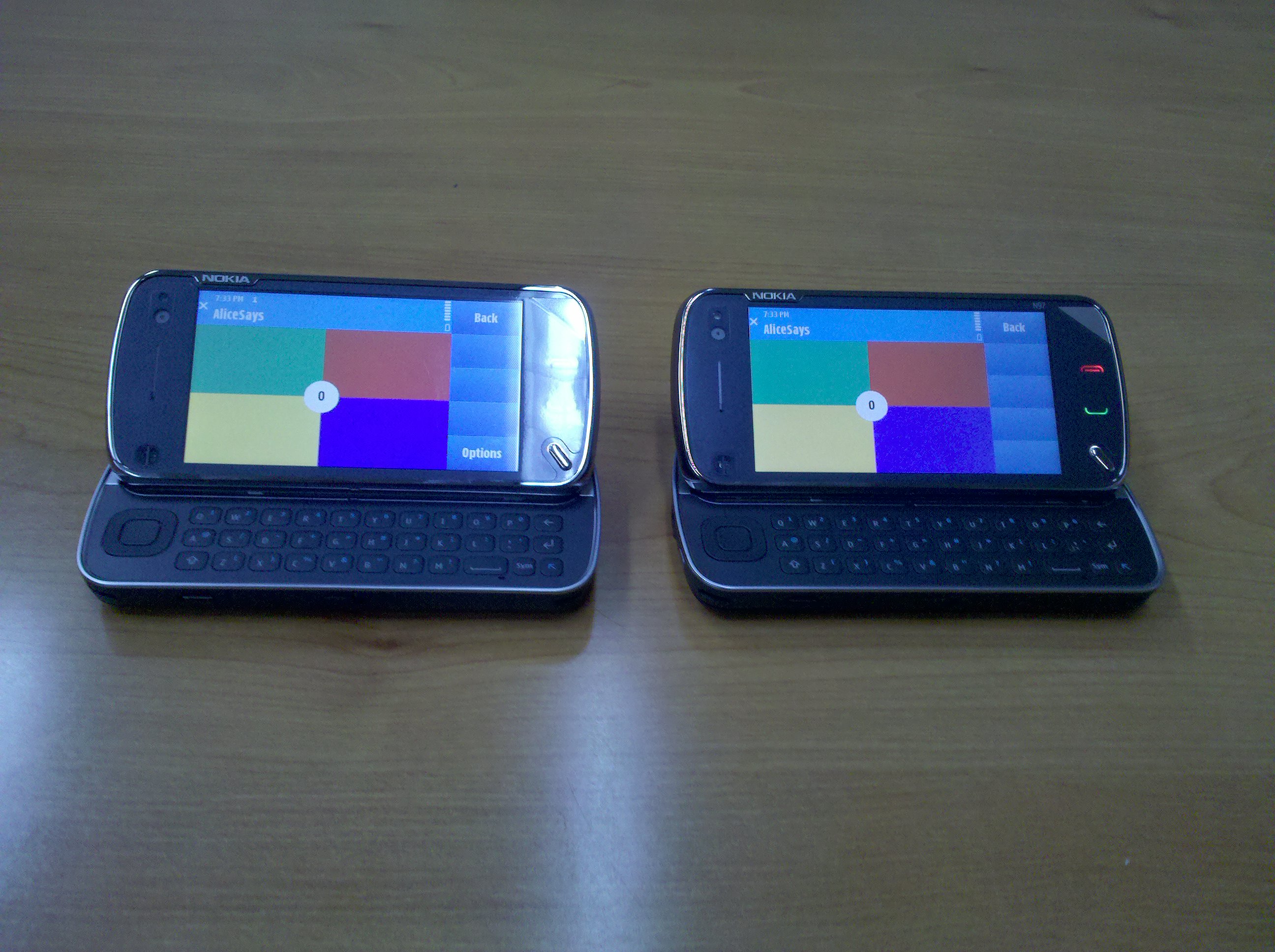}}
\caption{{Alice Says Game Set-up}}
  \label{fig:setup}
\end{figure}

\begin{figure}[htbp]
\centering{
  \includegraphics{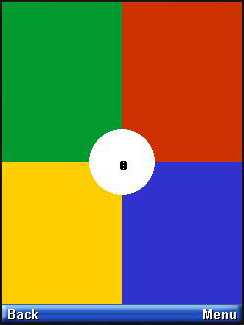}}
\caption{{Close Up Image of the Alice Says User Interface}}
  \label{fig:emulator_interface}
\end{figure}

The Nokia N97 smartphones provided numerous interfaces that were well suited
for exploring the usability of device pairing. For example, they featured a
resistive touch screen which allowed users to interact directly with what is
shown on its display. This imbued the game with a more realistic user
experience. Also enhancing the usability of the game was the stereo speakers
featured on the N97s, which are capable of outputting high quality sound even
at lower volume settings. This allowed us to keep the game's melodic tones as
close to those used in the original Simon game as possible. A picture of Alice
Says set-up from our implementation is shown in Figure \ref{fig:setup} and a clearer image of the game's core user interface, taken from an emulator, is displayed in Figure \ref{fig:emulator_interface}.

%\subsection{Other Implementation Details}

%move this to testing framework:
%It was important to decide which random strings to use for this mode, so we
%based our generation of random numbers on the default rand library of Java
%(java.util.Random). This created a simple and easy way to keep generating
%pseudo-random numbers every time the user made a mistake or beat the first 15
%levels (\texttt{Random rand = new Random(System.currentTimeMillis())}.

%\subsection{Role of the User}

\section{Usability Experimentation}

\label{sec:experimentation}

A user study was conducted in order to assess the viability of game-based
pairing in general and Alice Says in particular. These experiments were
intended to achieve several goals. Primarily, their focus was on gaining an
understanding of the usability of this pairing process. These tests were
intended to determine if the new game software could be used by real users to
consistently and reliably pair devices. In other words, we desired to discover
whether users felt that playing a game while pairing was useful or not. In
addition, a secondary test objective was to collect feedback from users regarding
Alice Says. A final aim was to ascertain the efficiency of Alice Says by
measuring its overall practical execution time.

\subsection{Experimental Framework}
\label{sec:framework}

In order to evaluate Alice Says, an environment was established where users
were able to get hands on experience with our prototype of the game. Review
\ref{sec:implementation} for full details on the implementation of this new
pairing system. Our intention was to create a testing framework that mirrored a
real world usage scenario as closely as possible while remaining uncomplicated.
An example of this is the omission of an in-band wireless link between the two
mobile devices. Since this process does not have a noticeable impact on the
usability of the pairing game, it was safely left out in favor of OOB strings
created using a pseudorandom pattern generated on a desktop machine using the
Mersenne Twister \cite{mersenne_twister} implementation included in the random
module of the Python programming language's standard library. The strings
utilized in these test cases were pseudorandomly generated, but fixed from
subject to subject to prevent some volunteers from receiving strings that were
easier to identify than others. These strings were presented to the test
subjects in a random order. This was done to minimize the effects of learning
and fatigue on the test results. In other words, we wanted to prevent users
from anticipating future test cases based on previous ones or losing motivation
to pay proper attention during the pairing procedure.

An automated feedback interface was used on a desktop computer to facilitate
this usability analysis. This was used to present post-condition questionnaires
to users following their completion of the study proper. In addition to the SUS scale, the exact phrasing of the post-condition survey questions that were presented to volunteers were as follows:

\begin{enumerate}
\item The method was enjoyable.
\item The method took a long time.
\item I would like to pair with another user's devices by making use of this method.
\item The sound effects used in this method were pleasant to listen to.
\end{enumerate}

Beyond this, logging was performed on the mobile devices themselves in order to capture both the timing of the events comprising the pairing procedure as well as any user mistakes that were made along the way.

\subsection{Participant Information}
\label{sec:participants}
%recruitment, incentives, where

20 test subjects were invited to participate in this usability survey. 
%This user threshold was selected due to its acceptance by the usability
%research community as the sample size required to capture over 98\% of
%usability issues as established by Faulkner \cite{20users}. 
Participants were gathered from students, professors, and staff members
studying and working in labs at our institution. Word of the study was spread
using flyers, emails, and in-person sign ups during the weeks prior to its
occurrence. To encourage participation in our study, movie theater gift
certificates were offered to testers upon the completion of their involvement
in the study. The post-condition survey questions themselves are provided in Section \ref{sec:framework}.

%% NS: Why are we guessing the age? Didn't we ask them? This needs to be cleaned up

Demographic information about these testers was collected as a component of the pre-conditioning phase of the experiment. Our subjects had varying ages, but were predominately young as would be expected from a pool containing university students. Half were between 18 and 24, while 30\% had ages of 25 to 29. There were several older individuals counted in our survey as well, however. 5\% of our sample population, or one user, was older than 29 but younger than 35. An additional user, comprising another 5\% of our user group, was between 35 and 39 years of age. Finally, 10\% of our sample was made up of people whose age exceeded 40. A majority, 65\%, of subjects were male although many females were also represented, comprising 35\% of the total. Another area in which sampling college students had an effect was on the level of education our testers had obtained. One tester (5\%) did not possess a college degree, 65\% had obtained their bachelor's, 25\% had obtained their master's degree, and one user (5\%) had completed his or her doctorate.

Beyond demographics, the post-conditional survey presented users with queries
intended to measure their level of expertise with device pairing and video
games. 70\% of participants had paired a wireless device before. This was not
somewhat surprising result considering the ubiquitous nature of wireless
devices and the need to pair them. This can perhaps be attributed to a
misunderstanding of what constitutes device pairing, as it could be anticipated
that most participants had utilized a wireless connection on their laptop or
paired a wireless remote control with a television or gaming appliance.
Remarkably, every participant responded that they played video games. This
figure demonstrates the widespread popularity of computer games and bodes well
for their acceptance by users as a means of facilitating secure usage habitats. 

\subsection{Experimental Design}
%try to make as reproducible as possible

%during
In this section, the design choices necessary to recreate the results of this
study are explored. To initiate the experiment, the test administrator first
selected Alice Says, represented by a Alice icon, from the mobile device's
application list. The single player version of Alice Says was then initiated by
selecting the ``1 Player'' option from the application's menu. Users were then
presented with the mobile device to provide them with an opportunity to
acclimate themselves with the user interface and overall gameplay of Alice
Says. 
%Most users picked up the game play naturally, but those who were confused
%were provided with a few brief instructions on how the game proceeded. 
Once
they felt that they had gained enough experience, users halted the single
player game section by selecting a back button from the game menu.

Next, the formal testing procedure was started by selecting complimentary game
modes on each of the two devices. This was done via a text field that
numerically specified one for the output device and two for the input device.
Matching test cases were specified on the two devices in an identical manner.
Once the options were put into place, the two player pairing mode was initiated
by selecting a ``2 Player'' choice from the menu on the mobile device. This
brought the primary Alice Says interface, consisting of four colored quadrants
and a central pattern score counter, up on their devices. As a final step to
initiate testing, the ``Start'' menu option was pressed on each device.

For the first two test cases performed with each subject, the participant
handled the input device while the test administrator took care of the output
duties. Thus, to start the game and whenever the volunteer successfully matched
a pattern, the administrator selected the ``Next'' option on the Alice Says
menu. In the event that the participant committed a mistake and did not
successfully match the provided pattern, the test conductor pressed a
``previous'' button to signal to the device that it should create a new pattern
starting with the last bit of the incorrectly matched pattern. Meanwhile, the
tester's job was to observe the audiovisual pattern displayed on the output
device and input it on their mobile appliance. Two test cases were performed in
this configuration to ensure that owe provided our subjects with system
experience that was generalizable and not specific to any individual test case.
Following this, a third and final test case was performed with the role of the
administrator and subject reversed by switching the mode options on the two
mobile phones. This was done to give users hands on experience with both sides
of the two player game; this is important because in a real life scenario this
will often be done to achieve mutual authentication as discussed in Section
\ref{sec:design}.

%post
After the conclusion of the central portion of the experiment, subjects were
presented with a web form containing a set of post-conditioning queries. Beyond
the demographic and background information listed in Section
\ref{sec:participants}, this questionnaire consisted of fifteen five point
Likert items which were selected to gauge how volunteers felt about Alice Says.
The precise questions posed are provided in Section \ref{sec:framework}. The
first ten of these questions were provided to evaluate this technique using the
System Usability Scale (SUS) \cite{sus}, modified slightly to refer to Alice
Says as a method rather than a system.

\subsection{Experimental Results}

%provide the raw data of the results label everything extremely carefully!
The observed results of the usability study are presented in this part of the
paper. Each test subject performed two input sessions and one output session
for a total of three test cases per user and 60 test cases overall. The average
time that users took to pair using Alice says was 173.267 seconds with a
standard deviation of 28.638 seconds. In terms of reliability, an average of
1.517 mistakes were made while performing the pairing process (i.e., an average
of 1.517 mistakes per pairing session).  Note that these are all partial errors
as pairing still completed successfully following their occurrence for all test
cases.  
%No simulation of attacked sessions was performed as part of this
%experiment, thus no fatal errors were observed. 
Some user errors were caused by
an inability to recall the displayed pattern, while others were caused by users
accidentally pressing the incorrect color button, though it is difficult to
separate these two occurrences in practice.

\subsubsection{User Feedback}

\begin{figure}[htbp]
\centering{
  \includegraphics[width=.45\textwidth]{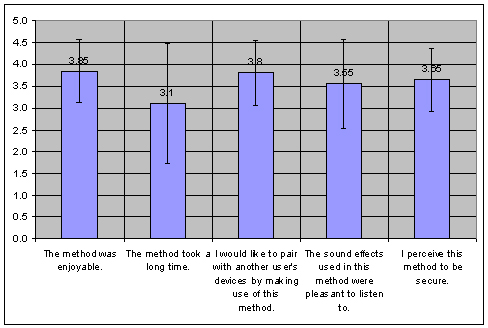}}
\caption{{{Average Responses to Post-Conditional Questionnaire}}}
  \label{fig:survey_results}
\end{figure}

Figure \ref{fig:survey_results} provides the responses users gave on the
post-condition survey. Users awarded Alice Says with an average SUS score of
70.5 and a standard deviation of 12.860. In response to whether or not Alice
Says was enjoyable, users provided an average response of 3.85. When asked if
this pairing process took a long time, users responded with a 3.1. However,
they also suggested that they would like to utilize Alice Says by providing the
highest average score observed, 3.88, to question thirteen. Users tended to
agree that the sounds used in the game were pleasant, providing a 3.55 average
response to this query. Finally, on the last question of whether Alice Says was
considered to be secure, users again replied positively with a 3.65.

Some users also responded to our open-ended question about Alice Says.  They
positively indicated that the paring method is ``pretty good'', ``a fun way to
pair'', ``a very cute game'', and ``a nice game'' and that ``they had a lot of
fun''.  However, some of them complained that method is ``too long''.  Some
suggested, as an improvement, to make the game ``less time consuming and
user-friendly'', use better menu options than ``next and previous'', make use
of ``better touchscreen'', and have the game ``count with me while I am picking
the colors''.

\subsection{Interpretation of Results}

This section outlines the implications of our test results.

\subsubsection{Efficiency}

The efficiency of Alice Says is clearly its least desirable characteristic. It
took slightly under three minutes to complete on average, when alternative
pairing approaches may take approximately twenty to thirty seconds (see
\cite{kstu09}, e.g.) for a similar level of security (i.e., corresponding to 30
bits). While this extended execution time would be problematic for pairing
approaches that burden users with tedious and repetitive tasks, this time frame
is less problematic for a game based pairing method. In a way, if a user is
having a good time while executing the pairing process, he or she may wish to
extend the technique's execution time rather than reduce it. However, contrary
to our intuition, users did indicate the game to be a bit lengthy. This is one
of the most important lessons learned from this study. In our future work, we
will experimentally determine the timing threshold that average users deem
appropriate for a pairing game and modify our game design to adhere to this
threshold (our current Alice Says prototype was not optimized in terms of
timing). 
%We believe that in order for the users to enjoy the overall process,  

\subsubsection{Reliability}

On the other hand, the reliability of Alice Says is one of its strengths. While
users committed, on an average, around one and half errors per session, the
pairing game was designed to be resilient in the face of such mistakes, and as
a result pairing concluded successfully in each and every attempt we performed.
We believe that as users become more or more familiar with this approach (and
even with the single player mode of the game), these errors will further be
reduced. This, in turn, would also reduce the overall execution timing of the
game, thus improving usability.

\subsubsection{User Feedback}

Considering that the industry average for SUS scores for computer systems are
somewhere between 60 and 70 \cite{sus-ratings}, our overall SUS score for Alice
Says can be termed quite positive. Looking at the questionnaire as a whole,
test subjects generally agreed with the positive statements regarding Alice
Says.  A majority of users concurred that the method was enjoyable, and seemed
secure. Furthermore, they also replied that they would like to make use of it.
Unfortunately, as discussed above, users also agreed with the one negative
statement that was put forth. This was in relation to the timing of the
technique, which was already known to be longer than users are accustomed to.
This aspect of the feedback prompts the need for further work on designing
games that are reasonably fast for the pairing process. 

The raw feedback provided by the users that they found the pairing process to
be entertaining was heartening and is promising towards the adoption of Alice
Says in practice.  

\subsection{Ecological Validity}

This section documents the ways in which this study conformed and deviated from
the real life situation it was intended to capture. The largest difference
between this study and an actual setting is the absence of a wireless link
between the two phones being paired. This may have had an impact on efficiency
because the latency this connection would introduce was not taken into account
in our study, but this step takes negligible time to execute compared to the
other steps involved in Alice Says. Beyond this, the only other ecological
concerns associated with this experiment are those encountered by all usability
studies in general. For example, by performing the tests in a lab rather than
in their own home, test subjects may have performed with a heightened awareness
of their actions. Testers may have altered their responses to survey questions
due to a desire to please the investigators, though this effect was guarded
against by providing subjects with privacy while answering as well as by
anonymizing the results. Finally, providing motivation for the test helped
users have a better understanding of how their actions related to real life
scenario, but also may have had a conditional effect. A pre-condition
questionnaire was not provided in order to minimize this impact, however.

\section{Discussion}

\label{sec:discussion}

\subsection{Lessons Learned and Improvements}

%% NS: We should clearly outline what are the lessons learned from our experience with Alice Says implementation and usability study and provide design guidelines for future games for pairing
%% NS: talk about things like sending the "result" bit embedded into game data, and mutual authentication via games pairing

The most important lesson learned via the usability study of Alice Says is the
importance of efficiency (speed) in the domain of device pairing. Prior to
conducting the usability experiments outlined in this paper, it was clear to us
that the execution time of our game prototype compared unfavorably to existing
solutions. We thought that due to the fact that users were enjoying themselves
while performing the process, the slowness of the process would not cause a
problem but rather help the users relish the game. This was also reflected from
the game-based solution for random number generation by Halprin and Naor
\cite{game_rng}, which did not suffer from any negative usability effects
despite taking far longer to complete (approximately 2 minutes to generate a
128-bit random number \cite{game_rng}) than traditional user input based
techniques.  Unfortunately, and surprisingly, this turned out not to be the
case for the scenario of pairing.  In this setting, it appears that users place
a high priority on speed.

Therefore, our results prompt the need for designing pairing games that also
aim at minimizing the execution time. To this end, one immediate task would be
to optimize Alice Says prototype in terms of speed. This can be done, e.g., by
determining an optimal duration for which individual colors are flashed
(currently this duration was set to be 300 ms which we thought was quite
convenient). Broadly speaking, future design of fast pairing games presents a
research challenge, however. This is because most existing games are not designed to be
completed in a matter of seconds, but rather take minutes or hours to complete,
as players wish to prolong the process to continue enjoying the experience.
However, there has been a movement towards very brief games played in rapid
succession, known as microgames, in the video game community that may be
adaptable as a solution.  Nintendo's WarioWare series is a prime example of
this \cite{warioware}.

Other possible improvements to Alice Says include adding another counter somewhere on
the user interface that ``counts with'' the user as they are inputting the
color pattern. Though this would be a stark deviation from the design of the
original game, it was a feature suggested by one of the test subjects as
something that would have a positive impact on the game's usability. Another
improvement suggested by users would be use devices with more responsive touch
screens. Several complaints were made about how the touch screen on the
Nokia N97s had a negative impact of the usability of the game as a whole.
Perhaps a device with a more modern capacitance based touch interface would not
suffer from this drawback. 

The efficiency problem would become even worse in cases where mutual
authentication is required; close to 6 minutes of pairing time would be needed
if our current Alice Says prototype is used. This may not be acceptable in
practice. The game playing part can be omitted in one of the directions by
having the user transfer the result of pairing from one device to the other (as
suggested in \cite{seka06}). However, this would make the pairing process
vulnerable to safe and fatal errors as well as prone to rushing user behavior,
similar to several other existing approaches (e.g., \cite{uka06}).  We note
that Alice Says can be effectively used to address this particular problem.
%of human errors in transmitting the result of pairing from one device to the
%other while mutual authentication is desired (as discussed above).  
For instance, the transmission of OOB string from device $A$ to $B$ could take
place via traditional means (such as using numeric representations), which will
be much faster, but the result ``bit'' can be transmitted via a game similar to
Alice Says by hiding the result bit within a short random string. This would
address both the problem of users being unmotivated and erratic while selecting
the correct option on device $A$, as well as that of the potential slowness of
the pairing process.

\subsection{Other Security Applications}

%% NS: Talk about authentication (e.g., mobile phone managers); or CATCHA's etc.

As future work, we intend to explore how the Tom Sawyer effect can be applied
to various security issues to enhance usability. As discussed in Section
\ref{sec:games-security}, Halprin and Naor \cite{game_rng} have already applied
this principle to the dilemma of random number generation to great effect.
Another area where it may be fruitful to apply this concept is that of
authentication in a variety of settings. One possibility is to improve the
usability of current mobile phone password managers (such as KeePassMobile
\cite{keepass}). Such phone managers suffer from poor usability in that the user
is required to manually transfer a (potentially long and random) password
displayed on the screen of the phone over to the authentication terminal. To
address this drawback, games similar to that of Alice Says can be adopted.

Another application is authentication to a remote server, basically as a
replacement for a CAPTCHA mechanism.  In order to prove to remote servers that
a human user is really behind a given request, users will be challenged to play
a game that is relatively easy for humans to complete but difficult for
computers. Finally, games may be designed to supplement the security and
privacy of ``something you have'' authentication techniques by having users
play a short movement game in order to unlock their access tokens, such as RFID
tags. This idea is similar in spirit to the recently proposed Secret Handshakes
scheme \cite{sh}, but it is aimed at providing an enhanced level of usability. 

\section{Conclusions}

In this paper, we considered the problem of designing pairing methods that incentivize users, in some way, so that they put forth more effect and correctly take part in the pairing process, thus providing improved security as well as enhancing the overall user experience. We dubbed this the Tom Sawyer Effect. To this end, we proposed a general direction of the application of computer games to solving tricky issues in usable security. The incentive that we provide to the users while they pair their devices is fun and entertainment. Since games are a popular means of entertainment, our hypothesis was that they may improve the security as well as usability of pairing and help solve the challenges outlined above.

We developed ``Alice Says,'' a pairing game based on a popular memory game called Simon (Says), and discussed the underlying design challenges. We also presented a \textit{preliminary} evaluation of Alice Says via a usability study and demonstrated its feasibility in terms of usability and security. Our results indicate that overall Alice Says was deemed a fun and an enjoyable way to pair devices, confirming our hypothesis. It was also found to be robust to human mistakes. However, contrary to our intuition, the relatively slower speed of Alice Says pairing was found to be a cause of concern, which prompts the need for the design of faster pairing games. We discussed some variations and possible improvements to our current Alice Says prototype, and more broadly, also outlined some of the other interesting applications of games to address lingering security problems.

%\end{document}  % This is where a 'short' article might terminate

%ACKNOWLEDGMENTS are optional
\section{Acknowledgments}
%This section is optional; it is a location for you
%to acknowledge grants, funding, editing assistance and
%what have you.  In the present case, for example, the
%authors would like to thank Gerald Murray of ACM for
%his help in codifying this \textit{Author's Guide}
%and the \textbf{.cls} and \textbf{.tex} files that it describes.

The authors would like to thank Frank Lantz for providing useful suggestions for research directions at the start of this project.

This work was partially supported by an NSF grant 0831397 and the United States
Department of Education GAANN grant P200A090157.

\bibliographystyle{abbrv}
\bibliography{soups}

\begin{thebibliography}{10}

\bibitem{simon}
{Simon}.
\newblock Available at \url{http://www.boardgamegeek.com/boardgame/5749/simon},
  1978.

\bibitem{simon_web}
{Flash Simon Game}.
\newblock Available at \url{http://www.thepcmanwebsite.com/media/simon/}, 2010.

\bibitem{Balfanz02}
D.~Balfanz, D.~Smetters, P.~Stewart, and H.~Wong.
\newblock {Talking To Strangers: Authentication in Ad-Hoc Wireless Networks}.
\newblock In {\em Network \& Distributed System Security Symposium (NDSS)},
  2002.

\bibitem{sus}
J.~Brooke.
\newblock {SUS - A quick and dirty usability scale}.
\newblock In {\em Usability Evaluation in Industry}, 1996.

\bibitem{keepass}
{C. Sperle}.
\newblock {KeePassMobile}.
\newblock Available at \url{http://www.keepassmobile.com}, 2010.

\bibitem{CCH06}
M.~Cagalj, S.~Capkun, and J.~Hubaux.
\newblock {Key agreement in peer-to-peer wireless networks}.
\newblock In {\em Proceedings of the IEEE}, 2006.

\bibitem{sh}
A.~Czeskis, K.~Koscher, J.~Smith, and T.~Kohno.
\newblock {RFIDs and Secret Handshakes: Defending Against Ghost-and-Leech
  Attacks and Unauthorized Reads with Context-Aware Communications}.
\newblock In {\em ACM Conference on Computer and Communications Security},
  2008.

\bibitem{MANA}
C.~Gehrmann, C.~Mitchell, and K.~Nyberg.
\newblock {Manual authentication for wireless devices}.
\newblock In {\em RSA CryptoBytes}, 2004.

\bibitem{Gold96}
I.~Goldberg.
\newblock {Visual Key Fingerprint Code}.
\newblock Available at {http://www.cs.berkeley.edu/iang/visprint.c}, 1996.

\bibitem{lac05}
M.~Goodrich, M.~Sirivianos, J.~Solis, G.~Tsudik, and E.~Uzun.
\newblock {Loud and Clear: Human-Verifiable Authentication Based on Audio}.
\newblock In {\em International Conference on Distributed Computing Systems
  (ICDCS)}, 2006.

\bibitem{kstu09}
A.~Kumar, N.~Saxena, G.~Tsudik, and E.~Uzun.
\newblock {Caveat Emptor: A Comparative Study of Secure Device Pairing
  Methods}.
\newblock In {\em International Conference on Pervasive Computing and
  Communications (PerCom)}, 2009.

\bibitem{2-user-tech}
A.~Kumar, N.~Saxena, and E.~Uzun.
\newblock {Alice Meets Bob: A Comparative Usability Study of Wireless Device
  Pairing Methods for a ``Two-User'' Setting}.
\newblock {\em CoRR}, 2009.

\bibitem{Kuo07}
C.~Kuo, J.~Walker, and A.~Perrig.
\newblock {Low-Cost Manufacturing, Usability, and Security: An Analysis of
  Bluetooth Simple Pairing and Wi-Fi Protected Setup}.
\newblock In {\em Usable Security (USEC)}, 2007.

\bibitem{gwap}
{L. von Ahn}.
\newblock {Games with a Purpose}.
\newblock In {\em IEEE Computer Magazine}, 2006.

\bibitem{reCAPTCHA}
{L. von Ahn, B. Maurer, C. McMillen, D. Abraham, M. Blum}.
\newblock {reCAPTCHA: Human-Based Character Recognition via Web Security
  Measures}.
\newblock In {\em Science Magazine}, 2008.

\bibitem{Nyberg05}
S.~Laur and K.~Nyberg.
\newblock {Efficient Mutual Data Authentication Using Manually Authenticated
  Strings}.
\newblock In {\em International Conference on Cryptology and Network Security
  (CANS)}, 2006.

\bibitem{sus-ratings}
J.~Lewis and J.~Sauro.
\newblock {The Factor Structure of the System Usability Scale}.
\newblock In {\em Proceedings of the 1st International Conference on Human
  Centered Design (HCD)}, 2009.

\bibitem{mersenne_twister}
{M. Matsumoto and T. Nishimura}.
\newblock {Mersenne Twister: A 623-Dimensionally Equidistributed Uniform
  Pseudo-Random Number Generator}.
\newblock In {\em ACM Transactions on Modeling and Computer Simulation}, 1998.

\bibitem{tomsawyer}
{M. Twain}.
\newblock {The Adventures of Tom Sawyer}, 1876.

\bibitem{MPR05}
J.~McCune, A.~Perrig, and M.~Reiter.
\newblock {Seeing-Is-Believing: Using Camera Phones for Human-Verifiable
  Authentication}.
\newblock In {\em IEEE Symposium on Security and Privacy}, 2005.

\bibitem{warioware}
{Nintendo}.
\newblock {WarioWare: Smooth Moves}.
\newblock Available at
  \url{http://www.nintendo.com/games/detail/7vgUzwrkjswZ6wiUXTtZQB8ji6_uPB3v},
  2010.

\bibitem{PV06}
S.~Pasini and S.~Vaudenay.
\newblock {SAS-Based Authenticated Key Agreement}.
\newblock In {\em International Conference on Theory and Practice of Public-Key
  Cryptography (PKC)}, 2006.

\bibitem{PS99}
A.~Perrig and D.~Song.
\newblock {Hash Visualization: a New Technique to improve Real-World Security}.
\newblock In {\em International Workshop on Cryptographic Techniques and
  E-Commerce (CrypTEC)}, 1999.

\bibitem{sr08}
R.~Prasad and N.~Saxena.
\newblock {Efficient Device Pairing using ``Human-Comparable'' Synchronized
  Audiovisual Patterns}.
\newblock In {\em Applied Cryptography and Network Security (ACNS)}, 2008.

\bibitem{game_rng}
{R. Halprin and M. Naor}.
\newblock {Games for Extracting Randomness}.
\newblock In {\em Symposium On Usable Privacy and Security}, 2009.

\bibitem{wisec}
V.~Roth, W.~Polak, E.~Rieffel, and T.~Turner.
\newblock {Simple and Effective Defense Against Evil Twin Access Points}.
\newblock In {\em ACM Conference on Wireless Network Security (WiSec)}, 2008.

\bibitem{seka06}
N.~Saxena, J.~Ekberg, K.~Kostiainen, and N.~Asokan.
\newblock {Secure Device Pairing based on a Visual Channel}.
\newblock In {\em IEEE Symposium on Security \& Privacy}, 2006.

\bibitem{su09}
N.~Saxena and M.~Uddin.
\newblock {Secure Pairing of ``Interface-Constrained'' Devices Resistant
  against Rushing User Behavior}.
\newblock In {\em Applied Cryptography and Network Security (ACNS)}, 2009.

\bibitem{suv08}
N.~Saxena, M.~Uddin, and J.~Voris.
\newblock {Universal Device Pairing Using an Auxiliary Device}.
\newblock In {\em Symposium On Usable Privacy and Security (SOUPS)}, 2008.

\bibitem{beda}
C.~Soriente, G.~Tsudik, and E.~Uzun.
\newblock {BEDA: Button-Enabled Device Association}.
\newblock In {\em International Workshop on Security for Spontaneous
  Interaction (IWSSI)}, 2007.

\bibitem{hapadep}
C.~Soriente, G.~Tsudik, and E.~Uzun.
\newblock {HAPADEP: Human Asisted Pure Audio Device Pairing}.
\newblock In {\em International Information Security Conference (ISC)}, 2008.

\bibitem{SA99}
F.~Stajano and R.~Anderson.
\newblock {The Resurrecting Duckling: Security Issues for Ad-hoc Wireless
  Networks}.
\newblock In {\em Security Protocols Workshop}, 1999.

\bibitem{uka06}
E.~Uzun, K.~Karvonen, and N.~Asokan.
\newblock {Usability Analysis of Secure Pairing Methods}.
\newblock In {\em Usable Security (USEC)}, 2007.

\bibitem{Vaudenay05}
S.~Vaudenay.
\newblock {Secure Communications over Insecure Channels Based on Short
  Authenticated Strings}.
\newblock In {\em CRYPTO}, 2005.

\end{thebibliography}
\end{document}